\title{Single source shortest paths in $H$-minor free graphs}

\author{
\em Raphael Yuster
\thanks{Department of Mathematics, University of Haifa, Haifa
31905, Israel. E--mail: raphy@math.haifa.ac.il}
}

\date{}

\documentclass [letterpaper,11pt]{article}
\usepackage{amsfonts}
\usepackage{graphicx}

\newtheorem{theo}{Theorem}[section]

\newtheorem{lemma}[theo]{Lemma}

\newcommand{\qed}{\hspace*{\fill} \rule{7pt}{7pt}}

\newcommand{\ignore}[1]{}

\begin{document}
\maketitle

\begin{abstract}

We present an algorithm for the Single Source Shortest Paths (SSSP) problem in \emph{$H$-minor free} graphs. For every fixed~$H$, if $G$
is a graph with $n$ vertices having integer edge lengths
and $s$ is a designated source vertex of $G$,
the algorithm runs in $\tilde{O}(n^{\sqrt{11.5}-2} \log L) \le
O(n^{1.392} \log L)$ time, where $L$ is the absolute value of the smallest edge length. The algorithm computes shortest paths and the distances from $s$ to all vertices of the graph, or else provides a certificate that $G$ is not $H$-minor free.
Our result improves an earlier $O(n^{1.5} \log L)$ time algorithm for this problem, which follows from a general SSSP algorithm of Goldberg.

\end{abstract}

\section{Introduction}
The {\em Single Source Shortest Paths} (SSSP) problem is the problem of finding shortest paths and, in particular, the distances from a specified source vertex to all vertices of a given directed graph. This problem is one of the classical and fundamental problems in computer science and has numerous applications.

Dijkstra \cite{Di-1959} gave an almost linear time algorithm
for the SSSP problem if all edge lengths are nonnegative reals. His algorithm runs in $O(m+n\log n)$ time where $m$
is the number of edges and $n$ is the number of vertices, if one uses the implementation from \cite{FrTa-1987}.
The situation becomes much more complicated when negative edge lengths are allowed.
Bellman \cite{Be-1958} and Ford \cite{FoFu-1962} gave an $O(mn)$ time algorithm for the SSSP problem where the edge lengths are arbitrary reals. No superlogarithmic improvement over this simple algorithm is known.
If the edge lengths are assumed to be integers, the fastest known algorithm to date is an $O(m\sqrt{n} \log L)$ of Goldberg \cite{Go-1993}, improving earlier algorithms of Gabow \cite{Ga-1985} and Gabow and Tarjan \cite{GaTa-1989}, where $L$ is the absolute value of the smallest edge length
(if $L$ is assumed to be a constant and $m$ is sufficiently large then there are slightly faster algorithms based upon fast matrix multiplication techniques).

A lot of research has been conducted in cases where the input graph belongs to some important family of graphs, as the SSSP problem has numerous applications even when the graph is a grid or a plane graph.
For sparse graphs with $m=O(n)$ edges, the above algorithm of Goldberg runs in
$O(n^{1.5}\log L)$ time. Unfortunately, even if we assume that the underlying graph has bounded degree, we do not know how to do better.
For planar graphs, however, better algorithms are known.
A strongly polynomial $O(n^{3/2})$ time algorithm was first given by Lipton, Rose, and Tarjan \cite{LiRoTa-1979}, based on the seminal
result on planar graph separators of Lipton and Tarjan \cite{LiTa-1979}.
For integer edge lengths, this was later improved by 
Henzinger et al. \cite{HeKlRaSu-1997} who gave an $\tilde{O}(n^{4/3}\log L)$ time algorithm.
As will be explained below, one {\em cannot} directly apply this algorithm to $H$-minor free graphs since finding an $O(\sqrt{n})$ separator for such graphs already requires $O(n^{1.5})$ time with present methods \cite{AlSeTh-1990}.
A significant improvement was made by Fakcharoenphol and Rao \cite{FaRa-2006} who gave an $O(n \log^3 n)$ algorithm for planar graphs with arbitrary real edge lengths.
Finally, Klein et al. improved this to $O(n \log^2 n)$ \cite{KlMoWe-2009}. These two algorithms are based on planarity.

In this paper we consider a much more general class of graphs, the class of $H$-minor free graphs.
A graph~$G'$ is a
\emph{minor} of a graph~$G$ if~$G'$ can be obtained from a subgraph of~$G$ by
contracting edges. A graph is \emph{$H$-minor free}
if~$H$ is not a minor of~$G$. In this paper, we say that a {\em directed} graph $G$
is $H$-minor free if the underlying undirected version of $G$ is $H$-minor free.
The classical Kuratowski-Wagner Theorem
\cite{Ku-1930,Wa-1937} states that a graph is planar if and
only if it has no $K_5$ nor $K_{3,3}$ minors. (For three different
proofs of the theorem, see \cite{Th-1981}.)

Families of $H$-minor free graphs, for some fixed
graph~$H$, are the cornerstone of the seminal theory of \emph{graph
minors} developed  over the last 20 years, in a series of more than
20 papers, by Robertson and Seymour. These families are, to date,
the most studied families of graphs in modern graph theory. The
graph minor theory of Robertson and Seymour culminated, in
\cite{RoSe-2004}, with a proof of the profound \emph{graph minor
theorem}, also known as the \emph{Wagner's conjecture}, that states
that in every infinite set of finite graphs, there is a graph which
is isomorphic to a minor of another. One of the consequences of this
theorem is that for any surface~$S$ (whether orientable or not) there is a finite set of
graphs~$F(S)$, such that a graph can be embedded in~$S$ (without
crossing edges) if and only if it does not contain a graph
from~$F(S)$ as a minor. (This result actually follows from a
restricted version of Wagner's conjecture which was already proved
in \cite{RoSe-1990}.) For a very recent survey of the theory of graph
minor see Lov{\'a}sz \cite{Lo-2006}.

Classes of $H$-minor free graphs are much more general, however,
than the class of planar graphs or classes of bounded genus graphs. For example, the class of
$K_5$-free graphs contains all the planar graphs, and many other
graphs, but there is no bounded genus surface on which all the
graphs from this family can be embedded.

The question we try to answer in this paper is the following:
Can we obtain an algorithm whose running time is significantly faster than
the $O(n^{1.5} \log L)$ time algorithm of Goldberg, when applied to $H$-minor free graphs (it is a well-known old result of Mader \cite{Ma-1967} that $H$-minor free graphs have $m=O(n)$ edges). Our main result yields a positive answer.
\begin{theo}
\label{theomain}
Let $H$ be any fixed graph. For a given $H$-minor free directed graph $G$
whose edge lengths are integers, and a designated source vertex $s \in V(G)$, there is an $\tilde{O}(n^{\sqrt{11.5}-2} \log L) \le O(n^{1.392} \log L)$ time
algorithm that computes a shortest path and the distance from $s$ to
each vertex of $G$.
\end{theo}
It should be noted that if $G$ contains an $H$-minor, the algorithm may still work as designated, but is allowed to fail. In case of failure, the algorithm produces a certificate for an $H$-minor in $G$.
An $H$-model in (the undirected version) of $G$ is a set of disjoint connected subgraphs $\{X_v ~:~ v \in V(H)\}$ indexed by the vertices of $H$, such that for every edge $uv \in E(H)$, there is an edge $xy \in E(G)$ with $x \in X_u$ and
$y \in X_v$. Clearly $G$ has an $H$-minor if and only if $G$ has an $H$-model.
Thus, in case of failure, the algorithm produces an $H$-model of $G$.

Our algorithm is based upon the ``four steps'' algorithm of Henzinger et al. \cite{HeKlRaSu-1997}, but with two major modifications, that are required due to the following obstacles.
It should be noted that \cite{HeKlRaSu-1997} works not only for planar graphs, but also for every graph which satisfies an $O(\sqrt{n})$-separator theorem (see the next section for an exact definition), {\em assuming} that an $O(\sqrt{n})$-separator can be obtained in linear time. Unfortunately, although $H$-minor free graphs satisfy
an $O(\sqrt{n})$-separator theorem, the fastest algorithm for finding such a separator, due to Alon et al. \cite{AlSeTh-1990} runs in $O(n^{1.5})$ time, and hence we cannot use it directly.
It has recently been shown by Reed and Wood \cite{ReWo-2005} that an
$O(n^{2/3})$-separator for $H$-minor free graphs can be found in
linear time. Using this result, it is already possible to obtain
an algorithm whose exponent is smaller than $1.5$, but not as good as the
exponent in Theorem \ref{theomain}. A bottleneck in the algorithm of \cite{HeKlRaSu-1997} is the need to solve many {\em all-pairs shortest paths} (APSP) problems in small, and very dense, pieces of the graph, and each such computation requires cubic time in the size of the pieces.
We show that by using separators that are slightly larger than optimal,
we can solve these APSP problems faster, resulting in a significant improvement in the overall running time. The main idea is
to first create the pieces quickly using large separators by using the
algorithm of \cite{ReWo-2005},
and within each small piece, use small separators, via the algorithm
of \cite{AlSeTh-1990}, to enhance the APSP computations.

We note that recently, Tazari and M\"uller-Hannemann \cite{TaMu-2008} obtained a linear $O(n)$ time algorithm for SSSP in $H$-minor free graphs where edge weights are {\em non-negative} reals,
extending another linear time algorithm of Henzinger et al. \cite{HeKlRaSu-1997} for planar graphs with arbitrary non-negative real weights.
The latter algorithm of Henzinger et al. (unlike their algorithm in the case of negative edge weights) works in linear time also for every bounded degree graph that has an $O(n^{1-\epsilon})$-separator theorem.
Thus, using the Reed-Wood result mentioned above, it directly works for $H$-minor free graphs with bounded degree. It is trivial to
transform any planar graph to another planar graph with bounded degree, while maintaining the shortest distances.
This, however, is far from trivial in the case of $H$-minor free graphs, and Tazari and M\"uller-Hannemann cleverly overcome this obstacle.

The rest of this paper is organized as follows. The next section contains definitions and lemmas that are needed for the proof of Theorem \ref{theomain}. The algorithm proving Theorem \ref{theomain} is described in Section 3. The final section contains some concluding remarks.
  
\section{Definitions and  Lemmas}

A {\em separation} of a graph $G$ is a pair $(A,B)$ of vertex sets
$A,B \subseteq V(G)$ such that $A \cup B = V(G)$, and there is no edge with one endpoint in $A \setminus B$ and the other
endpoint in $B \setminus A$. The set $A \cap B$ is called a
{\em separator} of $G$. Suppose now that $w : V(G) \rightarrow R^+$
is a weighting of $V(G)$. For $X \subset V(G)$ define
$w(X) = \sum_{v \in X} w(v)$ and define $w(G)=w(V(G))$.
We say that a graph $G$ with $n$ vertices has an
{\em $(f(n),\alpha)$-separator} if there is a separation $(A,B)$ with
$|A \cap B| \le f(n)$, $w(A \setminus B) \le \alpha w(G)$
and $w(B \setminus A) \le \alpha w(G)$.
We say that a hereditary family of graphs (a family closed under subgraphs) satisfies an {\em $(f(n),\alpha)$-separator theorem}
if every weighted graph with $n$ vertices belonging to the family has an {\em $(f(n),\alpha)$-separator}.

By the seminal result of Lipton and Tarjan \cite{LiTa-1979}, planar
graphs satisfy an $(O(\sqrt n), 2/3)$-separator theorem.
In fact, they also show how to compute an $(O(\sqrt n), 2/3)$-separator in linear time. Subsequently, Alon, Seymour and Thomas \cite{AlSeTh-1990}
extended the result of Lipton and Tarjan to $H$-minor free graphs.
The running time of their algorithm is $O(n^{1.5})$ for every fixed~$H$.

Clearly, if $g(n) \ge f(n)$ then having an $(f(n), \alpha)$-separator
implies having a $(g(n), \alpha)$-separator, but maybe the latter can be found more quickly. We thus say that a hereditary family of graphs
has an {\em $(f(n), \alpha, t(n))$-separator algorithm} if
it satisfies an  $(f(n), \alpha)$-separator theorem and an
$(f(n), \alpha)$-separator can be constructed in $O(t(n))$ time.
We can therefore state the result of \cite{AlSeTh-1990} as follows.
\begin{lemma}
\label{lemma-1}
For any fixed graph $H$, the family of $H$-minor free graphs
has an $(O(\sqrt n), 2/3, O(n^{1.5}))$-separator algorithm.
\end{lemma}

In a recent result, Reed and Wood \cite{ReWo-2005} generalize the
result from \cite{AlSeTh-1990} in an
interesting way. They show that a separator for an $H$-minor free
graph can be found more quickly, if we are willing to settle for a
larger separator. Quantifiably, their result can be stated as follows.

\begin{lemma}
\label{lemma-2}
Let $\gamma \in [0,1/2]$ be fixed and let~$H$ be
a fixed graph. The family of $H$-minor free graphs
has an $(O(n^{(2-\gamma)/3}), 2/3, O(n^{1+\gamma}))$-separator algorithm. Furthermore, if an input graph is not $H$-minor free
then an $H$-model asserting this fact is produced in $O(n^{1+\gamma})$ time. 
\end{lemma}
Notice that the case $\gamma=1/2$ of Lemma \ref{lemma-2} degenerates to
Lemma \ref{lemma-1}.

Suppose $G$ is a graph and $F \subset E(G)$. The {\em region} induced
by $F$ is the set of vertices incident with an edge of $F$.
A partition of $E(G)$ into $k$ parts defines a set of $k$ regions.
We say that a vertex of some region is a {\em boundary vertex}
if it belongs to more than one region. Otherwise, the vertex is
called {\em internal}.  
An {\em $(r,s)$-division} of an $n$-vertex graph $G$ is a partition of
$E(G)$ into $O(n/r)$ parts, so that each region contains at most $r$ vertices and $O(s)$ boundary vertices.

Fredrickson \cite{Fr-1987} showed that for every $r$, an 
$(r,\sqrt{r})$-division of an $n$-vertex planar graph can be found in $O(n \log n)$ time by a simple recursive application of the separator algorithm of Lipton and Tarjan. His method carries over without change 
to the more general setting of a hereditary family of graphs
with an $(f(n), \alpha, t(n))$-separator algorithm.
Thus, for $H$-minor free planar graphs we obtain, using Lemma \ref{lemma-2} and
\cite{Fr-1987}:
\begin{lemma}
\label{lemma-3}
Let $\gamma \in (0,1/2]$ be fixed and let~$H$ be
a fixed graph. For any $r \le n$, an $H$-minor free graph
with $n$ vertices has an $(r,r^{(2-\gamma)/3})$-division
and such a division can be constructed in $O(n^{1+\gamma}))$ time.
\end{lemma}

A family of sets $V_1,\ldots,V_k$ is called a {\em delta system}
if the common intersection of all of
them is identical to the intersection of any two of them.
\begin{lemma}
\label{lemma-4}
Let $G=(V,E)$ be a directed graph with $n$ vertices and with
$V=V_1 \cup \cdots \cup V_k$ where $V_1,\ldots,V_k$ is a delta
system of common intersection cardinality $t$, and $k$ is a constant.
Suppose that $G_i$ is the subgraph induced by $V_i$, and suppose that
an APSP solution for $G_i$ is given for all $i=1,\ldots,k$.
Then, an APSP solution for $G$ can be computed in
$O(n^2t)$ time.
\end{lemma}
{\bf Proof:}\,
We show how to compute the $n \times n$ distance matrix of $G$ given
the distance matrices of the $G_i$ in the claimed running time.
(The construction of the $n \times n$ predecessor matrix representing the shortest paths given the predecessor matrices of the $G_i$ can be computed similarly in the same time.)

Let $T = \cap_{i=1,\ldots,k} V_i$ and let $W_i = V_i \setminus T$.
Thus, $V=T \cup W_1 \cup \cdots \cup W_k$ is a partition of $V$.
Let $D_i$ denote the given distance matrix of $G_i$ for $i=1,\ldots,k$.
Thus, $D_i(u,v)=\delta_i(u,v)$ is the distance from $u$ to $v$ in $G_i$
where $u,v \in V_i$. Let $D$ denote the distance matrix of $G$.

We initially compute $D(u,v)$ in the case where both $u,v \in T$.
Define a complete directed graph $G_T$ on the vertices of $T$ by
setting the edge length of $(u,v)$ to be
$$
w_{G_T}(u,v) = \min_{i=1}^k \delta_i(u,v).
$$ 
As $k$ is constant and since the $D_i$ are given, the directed graph $G_T$
is constructed in $O(t^2)$ time.

Next, we solve the APSP problem in $G_T$ in $O(t^3)$ time using,
say, the Floyd-Warshall algorithm. Let $D_T$ be the resulting distance
matrix of $G_T$. We claim that $D(u,v) = D_T(u,v)$ for all $u,v \in T$.
Indeed, any shortest path from $u$ to $v$ in $G$ is constructed of
segments of shortest paths, where each segment is a shortest 
path in some $G_i$ from some vertex $x \in T$ to some vertex $y \in T$.
Since the length of this segment is (at least) the length of the {\em single edge} in $G_T$ from $x$ to $y$, the claim follows.

We next compute $D(u,v)$ where $u \in T$ and $v \in V \setminus T$.
Suppose $v \in W_i$. Either there is a shortest path in $G$ from $u$ to $v$ that is entirely contained in $G_i$, or else there is a shortest path
formed by the two segments from $u$ to some $z \in T$ and
from $z$ to $v$, where the latter is entirely contained in $G_i$. Notice that the distance of the first segment is
already computed as $D_T(u,z)$ and the distance of the second segment
is $D_i(z,v)$. It follows that
$$
D(u,v) = \min_{z \in T} D_T(u,z)+D_i(z,v).
$$
We therefore get that all the values $D(u,v)$ where $u \in T$ and $v \in V \setminus T$ can be computed in $O(nt^2)$ time.
Similarly, we compute $D(v,u)$ where $u \in T$ and $v \in V \setminus T$.

Finally, we compute the remaining $D(u,v)$ where $u,v \in V \setminus T$.
Either there is a shortest path from $u$ to $v$ that contains a
vertex of $T$, or else $u,v$ belong to the same $W_i$ and there
is a shortest path from $u$ to $v$ that is entirely contained in $G_i$.
Thus, setting $D'(u,v) = \min_{z \in T} D(u,z)+D(z,v)$
we have that, in case $u$ and $v$ are in distinct $W_i$
then $D(u,v) = D'(u,v)$. In case $D(u,v)$ are in the same $W_i$
then $D(u,v) = \min \{D_i(u,v)~,~ D'(u,v)\}$.
We therefore get that all the values $D(u,v)$ where $u,v \in V \setminus T$ can be computed in $O(n^2t)$ time.
\qed

\section{Proof of the main result}

Throughout this section we assume that $H$ is any fixed graph,
$G$ is an $n$-vertex directed graph with integer edge lengths, $s \in V(G)$ is a designated source vertex, and $-L$ is the smallest edge length appearing in $G$.
We show how to compute the distance from $s$ to each vertex of $G$ in the time stated in Theorem \ref{theomain}. The computation of the actual shortest paths (in the form of a predecessor tree) will be evident from the description.
We also assume that $G$ has no negative length cycles reachable from $s$. The algorithm can be easily modified to detect such a cycle if at least one exists.

Let $1/2 \ge \gamma > 0$ be a fixed parameter to be chosen later, and let $r$ be a function of $n$ to be chosen later. We follow the four steps algorithm from \cite{HeKlRaSu-1997} and apply the lemmas from the previous section
in the appropriate places.

\noindent
{\bf First step:}\,
We apply Lemma \ref{lemma-3} and obtain an
$(r,r^{(2-\gamma)/3})$-division of $G$ in $O(n^{1+\gamma})$ time.
We obtain a set ${\cal R}$ of $O(n/r)$ regions, where each $R \in {\cal R}$ has
$|R| \le r$ vertices, and has boundary $B(R)$ with $|B(R)| = O(r^{(2-\gamma)/3})$.

\noindent
{\bf Second step:}\,
The step applies the following procedure to each region
$R \in {\cal R}$. As in \cite{HeKlRaSu-1997}, the goal here is to obtain an auxiliary graph $H_R$ with the following properties.
The vertex set of $H_R$ is $B(R)$, and for each ordered pair of vertices $u,v \in
B(R)$ there is an edge $(u,v)$ in $H_R$ whose length is the distance from $u$ to
$v$ in $R$. In particular, $H_R$ is a complete directed graph (possibly with some edges having infinite length). However, the way we construct $H_R$ is
different from that in \cite{HeKlRaSu-1997} since we must avoid the na\"{\i}ve (say, Floyd-Warshall) application of APSP on a graph of size $O(r^{(2-\gamma)/3})$,
as this is too time consuming.

We will show how to create $H_R$ in $\tilde{O}(r^{11/6-2\gamma/3})$ time.
Using Lemma \ref{lemma-1} we find an $O(\sqrt{r})$ separator $X_R$ for $R$ that breaks $R$ into three pieces $T_1$, $T_2$, $T_3$ so that each piece contains a $2/3$-fraction of $R$ and a $2/3$-fraction of $B(R)$.
(by ``breaks'' we mean that after removing $X_R$ from $R$ the remaining vertices
are partitioned to $T_1$, $T_2$, and $T_3$).
Such a separator can be obtained by first finding an
$O(\sqrt{r})$-separator that breaks $R$ into two pieces, each containing at most a $2/3$-fraction of $R$, and then finding an
$O(\sqrt{r})$-separator of the part that contains more than half of the vertices of $B(R)$, so that no more than a $2/3$-fraction of the elements of $B(R)$ remain in a part after removing this second separator (note that here we use {\em weighted}
separators). 

Now define $R_i = T_i \cup X_R$ for $i=1,2,3$ and notice that the boundary of $R_i$
is (contained in) $X_R \cup (B(R) \cap T_i)$, and thus define $B(R_i)=X_R \cup (B(R) \cap T_i)$. Consider next each $R_i$ as a new (smaller) region, and 
apply the procedure recursively to obtain auxiliary graphs $H_{R_i}$
for $i=1,2,3$. How do we compute $H_R$, given the recursive computations of the $H_{R_i}$? Notice that the union of the $H_{R_i}$ is a delta system
with common intersection $X_R$ whose cardinality is $t=O(\sqrt{r})$.
Furthermore, the $H_{R_i}$'s are complete directed graphs and the edge lengths
of each $H_{R_i}$ constitute an APSP solution for the $H_{R_i}$.
Thus, an APSP solution for the union of the $H_{R_i}$ can be computed,
by lemma \ref{lemma-4} (with $n=O(r^{(2-\gamma)/3})$ in the statement of the lemma)
in time
\begin{equation}
\label{e1}
O(r^{1/2} \cdot (r^{(2-\gamma)/3})^2) = O(r^{11/6-2\gamma/3}).
\end{equation}
Since the union of the $H_{R_i}$'s contains $B(R)$ (and possibly some other vertices of $X_R \setminus B(R)$, we construct $H_R$ by setting, for all ordered pairs
$u,v \in B(R)$, the length of $(u,v)$ to be the distance from $u$ to $v$ in
the union of the $H_{R_i}$ (this distance is given to us from the APSP solution),
and this distance is clearly also the distance from $u$ to $v$ in $R$, as required.
Notice that the $O(\cdot)$ notation should be replaced with $\tilde{O}(\cdot)$ when summing up the running time over all the recursive calls, as this summation adds a logarithmic factor to the running time.
Thus, the overall time required to create the $H_R$ is $\tilde{O}(r^{11/6-2\gamma/3})$, as stated.
The time to create all the $H_R$ for all regions is, therefore,
$\tilde{O}(nr^{(5-4\gamma)/6})$.

\noindent
{\bf Third Step:}\,
We compute distances from $s$ to all the vertices in $\cup_{R \in {\cal R}} B(R)$. This is done by replacing each region $R$ with the complete directed graph $H_R$.
(Notice that an edge $(u,v)$ may have multiplicity now, since both $u$ and $v$
may appear together in more than one $B(R)$. Notice also that we assume that $s$ belongs to some $B(R)$ (when performing the first step, we have at least one region $R$ for which
$s \in R$, so we can artificially add $s$ to $B(R)$, if it is not already there).
The number of edges in the replaced graph (with multiplicities) is
$$
O\left(\frac{n}{r}(r^{(2-\gamma)/3})^2\right) = O(nr^{(1-2\gamma)/3}).
$$
The number of vertices in the replaced graph is $O(\frac{n}{r}r^{(2-\gamma)/3})$.
The smallest possible edge length in the replaced graph is trivially not smaller
than $-nL$. By applying Goldberg's $O(\sqrt{N}M \log K)$ single source shortest path algorithm for graphs with $N$ vertices, $M$ edges, and smallest weight $-K$
on our replaced graph, the distances from $s$ to all the vertices in $\cup_{r \in {\cal R}} B(R)$
can be computed in time
$$
\tilde{O}(n^{3/2}r^{(1-5\gamma)/6} \log L).
$$

\noindent
{\bf Fourth Step:}\,
We remain with the need to compute, for each region $R$,
the distance from $s$ to the vertices of $R \setminus B(R)$.
For this purpose, we construct an augmented graph $G_R$ which is obtained from $R$ by adding edges, and preserves the distances.
That is, if $u,v \in R$ then $\delta_R(u,v)=\delta_{G_R}(u,v)$.
However, the important feature of $G_R$ is that each distance can be obtained via a path of $G_R$ that has only $O(\log r)$ edges.
Cohen \cite{Co-1993} exhibits an efficient construction of such an augmentation,
but one can also use the construction described in the conference version of
\cite{HeKlRaSu-1997}. Indeed, in the second step, while recursively creating
$H_R$ from the $H_{R_i}$'s, we can also create $G_R$ recursively from $G_{R_i}$'s
be defining $G_R$ to be the union of $H_R$ and the $G_{R_i}$'s. An inductive argument shows that distances of $R$ are preserved in $G_R$, and that each distance in $G_R$ is obtained via a path consisting of $O(\log r)$ edges (corresponding to the depth of the recursion). Furthermore, the recursion shows that
$G_R$ has $O(r^{(4-2\gamma)/3} \log r)$ edges (indeed, recall that $H_R$ is a complete directed graph with $O(r^{(2-\gamma)/3})$ vertices).

Having constructed $G_R$, we obtain $G'_R$
by adding a new source vertex $s_R$ with an edge from $s_R$ to each vertex of $B(R)$
whose length is the distance from $s$ as computed in the third step.
Computing SSSP from $s_R$ in $G'_R$ using Bellman-Ford requires only
$O(\log r)=O(\log n)$ iterations, and each iteration is linear in the number of edges of $G'_R$ which is $O(r^{(4-2\gamma)/3} \log r)$.
Thus, shortest paths in $G'_R$ are computed in $\tilde{O}(r^{(4-2\gamma)/3})$ time
for each region, and in $\tilde{O}(nr^{(1-2\gamma)/3})$ time for all regions.
Clearly, the the computed distance from $s_R$ to a vertex $v \in R \setminus B(R)$ 
equals the distance from $s$ to $v$ in $G$.

Considering all four steps, the overall running time of the algorithm is
$$
\tilde{O}(
\max \{n^{1+\gamma}~,~ nr^{(5-4\gamma)/6} ~,~ n^{3/2}r^{(1-5\gamma)/6}\log L ~,~ nr^{(1-2\gamma)/3}\}) = 
$$
$$
\tilde{O}(\max \{n^{1+\gamma}~,~ nr^{(5-4\gamma)/6} ~,~ n^{3/2}r^{(1-5\gamma)/6} \log L \}). 
$$
For a given fixed $\gamma$, the optimal choice for $r$ is
$n^{3/(4+\gamma)}$ (in fact, it is  $n^{3/(4+\gamma)}(\log L)^{6/(4+\gamma)}$,
but we ignore this negligible improvement in the exponent of $\log L$)
which now means that the running time is
$$
\tilde{O}(\max \{n^{1+\gamma}~,~ n^{\frac{13-2\gamma}{8+2\gamma}}\log L \} ).
$$
Optimizing with $\gamma = \sqrt{11.5}-3 < 0.392$ the running time of the algorithm
is
$$
\tilde{O}(n^{\sqrt{11.5}-2} \log L)
\le O(n^{1.392}\log L)
$$
as required. \qed

It is interesting to note that the proof of Theorem \ref{theomain} yields a non-trivial complexity bound also when applied to {\em multiple sources}.
Indeed, the first and second step are not affected by the number of sources.
The third and fourth step can be applied to each source separately.
Thus, if we wish to compute distances from a set of $n^\alpha$ sources
to each vertex of the graph, the time required is
$$
\tilde{O}(
\max \{n^{1+\gamma}~,~ nr^{(5-4\gamma)/6} ~,~ n^{3/2+\alpha}r^{(1-5\gamma)/6}\log L ~,~ n^{1+\alpha}r^{(1-2\gamma)/3}\}).
$$
It is easy to see that for small values of $\alpha$ this is significantly
more beneficial than just performing the whole algorithm separately from each source, and is faster than any other presently known method.

\section{Concluding remarks and open problems}

Our algorithm for single source shortest paths in $H$-minor free directed graphs uses varying separator sizes,
and utilizes a tradeoff between the size of and the complexity of finding a separator. It would be interesting to find other applications of this method.
One such result, for the maximum matching problem, appears in \cite{YuZw-2007},
although in that result the size of the separator is not varying.

\end{document}